\documentstyle[12pt]{article}

\advance\textwidth 2.15cm
\advance\oddsidemargin -1cm
\advance\textheight 1.5cm
\advance\topmargin -1cm

\newcommand\beq{\begin{equation}}
\newcommand\eeq{\end{equation}}
\def\beqa{\begin{eqnarray}}
\def\eeqa{\end{eqnarray}}
\def\bega{\begin{array}}
\def\enda{\end{array}}
\def\be{\[}
\def\ee{\]}

\def\sp{\hspace{.8cm}}

\def\bq{{\bar q}}
\def\vb{\vphantom{b}}
\def\q#1#2{ { q^{#1\vb}_{#2} } }
\def\p#1#2{{p^{#1\vb}_{#2}}}
\def\hq#1#2{{q^{#1\vb}_{#2}}}
\def\hp#1#2{{p^{#1\vb}_{#2}}}

\def\sH#1{{\cal H}^{(#1)}}
\def\H#1{H^{(#1)}}

\def\c#1{_{,#1} }  

\def\m#1{{\mu^{(#1)}}}
\def\phc#1{\phi^{(#1)*}}
\def\ph#1{\phi^{(#1)}}
\def\ps#1{\psi^{(#1)}}

\def\Schrodinger{Schr\"odinger}

\def\a{\alpha}
\def\b{\beta}
\def\ga{\gamma}
\def\eps{\epsilon}
\def\la{\lambda}
\def\w{\omega}

\def\d{\partial}
\begin{document}

\title{Canonical Transformations in Quantum Mechanics}
\author{Arlen Anderson\thanks{arley@physics.mcgill.ca}
\\Department of Physics\\ McGill University\\ Ernest
Rutherford Building\\
Montr\'eal PQ Canada H3A 2T8 }
\date{May 20, 1992}
\maketitle

\vspace{-10cm}
\hfill McGill 92-29

\hfill hep-th/9205080
\vspace{10cm}

\begin{abstract}
Three elementary canonical transformations are shown both
to have quantum implementations as finite transformations and
to generate, classically and infinitesimally, the full canonical
algebra.
A general canonical transformation
can, in principle, be realized quantum mechanically as a product of
these transformations. It is found that the intertwining of two
super-Hamiltonians is equivalent to there being a canonical transformation
between them.
A consequence is that the procedure for
solving a differential
equation can be viewed as a sequence of elementary
canonical transformations trivializing the super-Hamiltonian associated
to the equation.
It is proposed that the quantum integrability of a
system is equivalent to the existence of such a sequence.
\end{abstract}
PACS: 03.65.Ca, 03.65.Ge
\newpage

Canonical transformations are a powerful tool of classical mechanics whose
strength has not been fully realized in quantum mechanics. Despite work by
Dirac\cite{Dir} and others\cite{Itz,Bar,Mos,qo}, there is a
persistent distrust of
the utility of canonical transformations in quantum mechanics. The general
feeling  lies between believing they can't be used in quantum
mechanics and thinking that at best they are of only limited interest. The
truth, however, is that effectively all exactly solvable problems in
quantum mechanics are, or can be, solved by canonical transformations. Each
of the familiar tools for integrating a differential equation---change of
variables, extracting a factor from the dependent variable, Fourier
transform, etc.---has a realization as a canonical transformation. Only
procedures whose origins lie in approximation, such as power series
expansion, are not essentially canonical transformations.

By making the quantum implementation of canonical transformations explicit,
the mechanics of solving differential equations is made more transparent.
This is especially useful in problems of more than one variable where
notational complications begin to cloud insight. Furthermore, the
usefulness of canonical transformations is not limited to exactly solvable
problems; they can also be important in preparing a problem for
approximation. As well, by bringing to quantum mechanics a tool so
central to our understanding of classical mechanics, our grasp of
their relationship is deepened, and new lines of investigation are opened.

Perhaps most significantly, canonical transformations provide a unifying
structure in which to study quantum integrability.  The method of
intertwining\cite{And}, in
which one constructs an operator $D$ which transforms between two operators
\beq
\label{ior1}
\sH1 D= D\sH0
\eeq
so that their eigenfunctions are related
\beq
\ps1=D\ps0,
\eeq
has already been proposed for this role\cite{An2}.  This is
because it underlies the factorization method of Infeld and Hull\cite{InH},
supersymmetry in quantum mechanics\cite{susy}, and Lie algebraic
methods\cite{lam}, which
together solve most integrable problems in quantum mechanics.  The method
of intertwining was limited in the past by the need to make an
ansatz for the intertwining operator $D$.  We shall see that the intertwining
operator is simply the canonical transformation between
$\sH0$ and $\sH1$.

This observation considerably deepens one's intuition for
constructing intertwining operators.  In addition, it explains why
intertwining underlies the  Lie algebraic methods\cite{An3}:  the symmetry
group
preserving the form of a differential equation is a subgroup of the full
canonical group.  More, the infinite-dimensional Kac-Moody algebra
associated,
for example, with the soliton solutions of the KdV equation, is
explained through the Lax formalism because it is a subalgebra
of the full canonical algebra.

In the following, three elementary canonical transformations and their
quantum implementations will be introduced.  Using these, several
additional composite elementary transformations will be constructed.  It
will be shown that, classically, the infinitesimal forms of these
transformations generate the full algebra of canonical transformations.
This implies that in principle we have the tools to make any quantum
canonical transformation.

In practice, we are limited to composing
a certain collection of finite transformations.  These transformations will be
recognized as familiar tools for solving differential equations:
changes of variable, similarity (gauge) transformations, Fourier
transform.  In addition, the standard first order differential
intertwining operator will be
realized as a canonical transformation.  It is noteworthy because it is a
transformation whose action is simpler in the quantum context than in
the classical.  Finally, a few examples are done to illustrate various
aspects of the use of canonical transformations.  Many more complicated
examples follow straightforwardly by use of the differential
intertwining canonical transformation.

As this paper was being prepared, the author became aware of recent work
by Leyvraz and Seligman \cite{LeS} and Deenen \cite{Dee} which overlaps
in part with the results given here.

\section{Formal aspects}

\subsection{Introduction}

The canonical transformations discussed in this paper may be applied to any
wave equation. For convenience, attention will be focused on the
non-relativistic  \Schrodinger\ equation. The presumption is that the
wavefunctions of an initial equation are of interest, and
the intent is to construct them by canonically transforming to simpler
equations.

Since canonical transformations involving the time $\q{}0$ and its
conjugate momentum $\p{}0$ are useful, it is profitable to work in the
extended phase space in which these are adjoined to the spatial coordinates
and momenta\cite{Lan}.
The \Schrodinger\ operator is the
super-Hamiltonian
\beq
{\cal H}({\bf \hq{}{},\hp{}{}}) =\hp{}0+H(\hq{}0,\hq{}1,\hp{}1),
\eeq
where ${\bf \hq{}{},\hp{}{} }$ are used to denote all of the
extended phase space coordinates.  For notational convenience, only one
spatial degree of freedom is indicated explicitly.
The  \Schrodinger\ equation in the coordinate representation
is the constraint equation
\beq
{\cal H}\psi({\bf q})=0.
\eeq

A canonical transformation is performed by applying an operator, say $P$,
on the left of the super-Hamiltonian constraint equation and
rewriting the result as a new constraint equation
\beq
0=P\sH0\ps0=\sH1\ps1.
\eeq
This gives the transformed super-Hamiltonian
\beq
\label{sHr1}
\sH1({\bf \hq{}{},\hp{}{} }) = P \sH0({\bf \hq{}{},\hp{}{}}) P^{-1},
\eeq
and the transformed wavefunction
\beq
\ps1=P\ps0.
\eeq

The operators $P$, $P^{-1}$ can be brought inside the super-Hamiltonian
to act directly on the phase space coordinates
\beq
\sH1({\bf \hq{}{},\hp{}{} }) =  \sH0(P{\bf \hq{}{}} P^{-1},P {\bf \hp{}{}}
 P^{-1}).
\eeq
This is possible
for any function on phase space possessing a Laurent expansion about some
point, as can be seen by expanding the
function, applying the transformation and resumming.
In practice it holds for all super-Hamiltonians of interest.
Since the canonical commutation relations are preserved by this
transformation
\beq
[{\bf q},{\bf p}]=[P{\bf q} P^{-1}, P{\bf p}P^{-1}]=i,
\eeq
the effect is to transform the original phase space coordinates by
the canonical transformation, which was the object.
Note that (\ref{sHr1}) is the same as the intertwining condition (\ref{ior1}).
This establishes the equivalence of the canonical transformation $P$
and the intertwining operator $D$.

The goal of constructing the wavefunctions of $\sH0$ is attained if, by a
sequence of canonical transformations, one can
transform to an $\sH1$ which is known to be integrable.
It is conjectured that a
super-Hamiltonian is integrable if and only if it can be canonically
transformed to the
trivial super-Hamiltonian
\beq
{\cal H}=\p{}0.
\eeq
The canonical transformation to a trivial super-Hamiltonian has special
significance in classical
mechanics:  this transformation makes all the variables constants of the
motion, and these constants may be chosen to be the initial conditions of
the system.  Inverting this transformation classically solves the
equations of motion.

The inner product is assumed to be conserved
\beq
\int dq\, \m0 \phc0 \ps0 = \int dq\, \m1 \phc1 \ps1
\eeq
(where $q$ runs over a spatial hypersurface\cite{time}).
This determines a transformed measure density, which in turn may be used to
define the adjoint of $P$.  Under this definition, $P^{\dagger}P=1$,
and $P$ is seen to be a unitary transformation.
In the general case, the measure density may be operator-valued---consider,
for example, the Wronskian operator appearing in the Klein-Gordon inner
product.  Details of the transformation of the measure density will not
be discussed in
this paper; they are straightforward to work out from the inner
product.

\subsection{Infinitesimal Canonical Transformations}

Classically, an infinitesimal canonical transformation is generated by an
(infinitesimal) generating function $F(q,p)$\cite{Gol}.  Associated to this
generating function is the Hamiltonian vector field
\beq
v_F=F\c{p}\d_q- F\c{q} \d_p
\eeq
whose action on a function $u$ on phase space is the infinitesimal
transformation
\beq
\delta_F u=\eps v_F u= -\eps \{F, u\}
\eeq
where $\{F,u\}$ is the classical Poisson bracket. Through the correspondence
between classical and quantum theory in which Poisson brackets go over
into commutator brackets times $i$, this canonical transformation can be
expressed in terms of the quantum operator $\exp(-i\eps \hat F)$
\beq
e^{-i\eps \hat F}\hat u e^{i\eps\hat F}= \hat u -i\eps [\hat F,\hat u]
+O(\eps^2).
\eeq
Because of operator ordering ambiguities in defining the quantum versions of
$\hat F$ and $\hat u$, the classical
and quantum
expressions for the infinitesimal transformation can
differ by higher order terms in $\hbar$ (at the same order of $\eps$).

In the equations that follow, quantum variables are not distinguished
notationally from classical ones.  Unless otherwise stated,
all variables are handled according to their familiar quantum
contexts, e.g. as
operators in Hamiltonians and as c-numbers in wavefunctions.
Products of operators are assumed to
be ordered as
written; classical versions are obtained by allowing $p$ and $q$ to
commute.

There are three elementary canonical transformations in one-variable which
have well-known
implementations as finite quantum transformations. They are similarity
(gauge) transformations, point canonical (coordinate) transformations and
the interchange of coordinates and momenta.  By finding the algebra
generated by the
infinitesimal versions of these transformations, the class of
transformations that can be reached with them can be determined. The
somewhat surprising result is that classically they generate the
full canonical
algebra.

The significance of this classical result is that since each of the
elementary and
composite elementary transformations has
a quantum implementation as a finite transformation, in principle any
general
canonical transformation can be implemented quantum mechanically.  In
practice at present, one is limited to finite products of the
transformations.  Nevertheless this is a very powerful tool for solving
problems in
quantum mechanics.

Similarity (gauge) transformations are
infinitesimally generated by  $F_S=f(\hq{}{})$.
Point canonical transformations are infinitesimally generated by
$F_P=f(\hq{}{})\hp{}{}$.  The discrete transformation $I$
interchanging the role of coordinate and momentum is
\beq
\hp{}{}=I\p{\prime}{}I^{-1}=-\hq{\prime}{}, \sp
\hq{}{}=I\q{\prime}{}I^{-1}=\hp{\prime}{}.
\eeq
Using the interchange operator, composite elementary transformations which
are nonlinear in the momentum can be formed.  They are the composite
similarity transformation, infinitesimally generated by
$$ F_{CS}=I F_S I^{-1} =f(\hp{}{}),$$
and the composite point canonical transformation, infinitesimally
generated by
$$F_{CP}=I F_P I^{-1} =-f(\hp{}{})\hq{}{}.$$
Each of these corresponds to a finite transformation through application
of the interchange operator to the finite forms of the similarity and
point canonical transformations.

The many-variable generalization of the similarity transformation is
straightforward.  It
is infinitesimally generated by $F_{Sn}=f(\hq{}1,\ldots,\hq{}n)$.
Interchanging  any set of coordinates with their conjugate momenta
gives the composite many-variable similarity transformations.  For example,
in two
variables, one has the infinitesimal generating functions
$F_{CS2a}=f(\hp{}1,\hq{}2)$ and $F_{CS2b}=f(\hp{}1,\hp{}2)$.

The observation is now made that classically the algebra
generated by
the elementary and composite
elementary transformations is the full canonical algebra
\beq
[v_F,v_G]=-v_{\{F,G\} }.
\eeq
where $F,G$ are arbitrary functions on phase space.  It is to be
expected that
Hamiltonian vector fields will have this algebraic structure.   More
surprising is that the above collection of generating functions
produce a general function on phase space through the Poisson
bracket operation.  This is verified by
taking commutators of the different types of transformations.

Consider the two-variable case---the many-variable case follows similarly.
Introducing the monomial generating functions
\beq
F^{jk}_{nm} = \q{\,j+1}{1}\q{\,k+1}{2}\p{\,n+1}{1}\p{\,m+1}{2},
\eeq
where $j,k,n,m\in Z$,
the Poisson bracket of two of these is
\beqa
\{F^{j_1 k_1}_{n_1 m_1},F^{j_2 k_2}_{n_2 m_2}\}&=&
\bigl( (j_2+1)(n_1+1)- (j_1+1)(n_2+1)\bigr)
F^{j_1+j_2\ k_1+k_2+1}_{n_1+n_2\ m_1+
m_2 +1} \\
&&+\bigl( (k_2+1)(m_1+1)-(k_1+1)(m_2+1) \bigr)
F^{j_1+j_2+1\ k_1+k_2}_{n_1+n_2+1\ m_1+m_2}. \nonumber
\eeqa
Inspection shows that a general monomial can be constructed by
beginning with the monomial forms
$F^{j\ -1}_{0\ -1}$, $F^{j\ k}_{-1 \ -1}$,
$F^{0\ -1}_{n\ -1}$, $F^{-1\ -1}_{n\ m}$, etc., which generate
the elementary
and composite elementary canonical transformations.

By taking linear
combinations, one can form any function having a Laurent expansion about
some point (not necessarily the origin).  To avoid having to use formal
Laurent expansions about nonzero points for functions
like $q^{1/2}$ or $\ln q$,
greater generality is
obtained by working with generating functions of the form
\beq
F=f_1(\q{}1) f_2(\q{}2) g(\p{}1) g(\p{}2).
\eeq
This produces any function which can be represented as a sum of separable
products.

\subsection{Quantum Implementations}

Each of the elementary canonical transformations can be implemented
quantum mechanically as a finite transformation.  Their action is
collected in Fig.~1, and each will be reviewed below.

The interchange of coordinates and momenta
\beq
\hp{}{}=I\p{\prime}{}I^{-1}=-\hq{\prime}{}, \sp
\hq{}{}=I\q{\prime}{}I^{-1}=\hp{\prime}{}.
\eeq
is implemented through
the Fourier transform operator
\beq
I={1\over (2\pi)^{1/2}} \int_{-\infty}^\infty dq e^{iq'q}
\eeq
for which it is evident that
\beq
I \hq{}{} =-i\d_{q'}I, \hspace{.8cm} I \hp{}{}  =-q'I.
\eeq
The wavefunction is transformed
\beq
\ps1(\q{\prime}{})=I\ps0(\q{}{})={1\over (2\pi)^{1/2}}
\int_{-\infty}^\infty dq e^{iq'q} \ps0(\q{}{}).
\eeq
The inverse interchange is
\beq
\hp{}{}=\hq{\prime}{}, \sp \hq{}{}=-\hp{\prime}{}.
\eeq
It is implemented by the inverse Fourier transform
\beq
I^{-1}={1\over (2\pi)^{1/2}} \int_{-\infty}^\infty dq e^{-iq'q}.
\eeq

The similarity transformation in one-variable is implemented by the operator
\beq
S_{f(\q{}{})}=e^{-f(\hq{}{})}
\eeq
where $f(\hq{}{})$ is an arbitrary function of the coordinates. This
induces the canonical transformation
\beqa
\label{s}
\hp{}{} &=& S_{f(\q{\prime}{})} \hp{\prime}{}
S_{f(\q{\prime}{})}^{{\vb}^{-1}} =\hp{\prime}{} -i
f\c{\hq{\prime}{}}, \\
\hq{}{}&=& S_{f(\q{\prime}{})} \hq{\prime}{}
S_{f(\q{\prime}{})}^{{\vb}^{-1}} = \hq{\prime}{} .
 \nonumber
\eeqa
The composite similarity transformation is implemented by applying the
interchange operator to $S_{f(\q{}{})}$ to exchange coordinates for
momentum in
$f$.  In the one-variable case, the composite similarity transformation
operator is
\beq
S_{f(\p{}{})}=I S_{f(\q{}{})} I^{-1} =e^{-f(\hp{}{})}.
\eeq
It produces the canonical transformation
\beqa
\label{cs}
\hp{}{} &=& S_{f(\p{\prime}{})} \hp{\prime}{}
S_{f(\p{\prime}{})}^{{\vb}^{-1}}
=\hp{\prime}{} , \\
\hq{}{}&=& S_{f(\p{\prime}{})} \hq{\prime}{}
S_{f(\p{\prime}{})}^{{\vb}^{-1}}
= \hq{\prime}{} +i
f\c{\hp{\prime}{}} . \nonumber
\eeqa

In the many-variable case, the function $f$ may involve either the
coordinate or its conjugate momentum for each variable.  Because the different
variables commute amongst themselves, each responds to the (composite)
similarity operator as if
it were a one-variable operator in that variable, and the other variables
are treated as parameters. Thus,
for each coordinate (momentum) of which $f$ is a
function, the corresponding conjugate momentum (coordinate) is shifted as
in the one-variable case.

The finite point canonical transformation takes a different form than the
infinitesimal version.  The effect of the point canonical transformation
$P_{f(\q{}{})}$
is to implement the change of variables
\beqa
\label{pc}
\hq{}{} &=& P_{f(\q{\prime}{})} \hq{\prime}{}
P_{f(\q{\prime}{})}^{{\vb}^{-1}}
=f(\hq{\prime}{}), \\
\hp{}{} &=& P_{f(\q{\prime}{})} \hp{\prime}{}
P_{f(\q{\prime}{})}^{{\vb}^{-1}}
={1\over f\c{\hq{\prime}{}} }
\hp{\prime}{}. \nonumber
\eeqa
The effect of $P_{f(\q{}{})}$ on the wavefunction is
\beq
\ps1(q)=P_{f(\q{}{})} \ps0(q)=\ps0(f(q)).
\eeq

The composite point canonical transformation is formed by composition
with the interchange operator
\beq
P_{f(\p{}{})}=I P_{f(\q{}{})} I^{-1}.
\eeq
It has the effect of making a change of variables on the momentum
\beqa
\hq{}{} &=&{1\over f\c{\hp{\prime}{} }}\hq{\prime}{},  \\
\hp{}{} &=& f(\hp{\prime}{}) . \nonumber
\eeqa
The operator ordering of the transformed $\hq{}{}$ is determined by
the action of the interchange operator on the coordinate point canonical
transformation.

A two-variable generalization is also of interest.  It
is defined by the canonical transformation
\beqa
\label{pc2}
\hq{}{1} &=& f(\hq{\prime}{1},\hq{\prime}{2}), \\
\hp{}{1} &=& {1\over f\c{\hq{\prime}{1}} }
\hp{\prime}{1}, \nonumber\\
\hq{}{2} &=& \hq{\prime}{2} , \nonumber \\
\hp{}{2} &=& \hp{\prime}{2}+
{f\c{\hq{\prime}{2}} \over f\c{\hq{\prime}{1}} }
\hp{\prime}{1}. \nonumber
\eeqa

Using this, the result of exponentiating
the infinitesimal
generating function $ F=g(\hq{}2)h(\hq{}1)\hp{}1$ can be deduced.  Since
different variables commute, the action of $e^{-i F}$ on $\hp{}2$ is
that of a
similarity transformation (\ref{s}) and we have
\be
e^{-i F}\hp{\prime}{2} e^{i F} =\hp{\prime}{2}+
 F\c{\hq{\prime}{2} }.
\ee
Equating this with the transformation in (\ref{pc2}) gives
\beq
{f\c{\hq{\prime}{2}} \over f\c{\hq{\prime}{1}}}=
g(\hq{\prime}2)\c{\hq{\prime}{2} }
h(\hq{\prime}1).
\eeq
Defining
\be
H(q)=\int {1\over h(q)} dq,
\ee
the chain rule simplifies this equation to
\be
f\c{H}-f\c{g}=0
\ee
which has the general solution
\be
f=f(H(\hq{}1)+g(\hq{}2)).
\ee
Since the transformation must reduce to the identity when $g=0$, we find
\beq
\label{pcf1}
f=H^{-1}(H(\hq{}1)+g(\hq{}2)).
\eeq
This agrees with a result found independently by Deenen.\cite{Dee}

As an explicit example, take $g={\rm const}$ and $h=\hq{\,m}1$.  The
infinitesimal
transformation is then that of the Virasoro generators.  The finite
transformation is (for $m\ne 1$)
\beq
f=(\hq{\,1-m}1 + (1-m)g)^{1/1-m}.
\eeq
while for $m=1$
\beq
f=e^g\hq{}1.
\eeq
I remark that for $m=0,1,2$ these transformations generate
the group $SL(2,C)$.

In the one-variable case with $g$ taken to be a
constant, Eq.~(\ref{pcf1}) gives the finite transformation produced by
the infinitesimal generating function $ F=g h(\hq{}{})\hp{}{}$.
It can also be read as a functional equation
\beq
H(f(\hq{}{}))=H(\hq{}{})+g
\eeq
which is to be solved for $H$ given $f$.  Not surprisingly, this equation
does not have a solution for all $f$.  The implication is that not all
point canonical transformations can be expressed as the exponential of an
infinitesimal transformation.  This is an explicit demonstration of the
well-known property of the diffeomorphism group that the exponential map
does not cover a neighborhood of the identity\cite{Mil}.  This is a
property of infinite-dimensional Lie groups and stands in contrast to
the situation in finite-dimensional Lie groups.

A corollary is that the product of the exponentials of two
generators cannot
always be expressed as an exponential of a third generator.  For this
reason, it is generally not useful to express point canonical
transformations in exponential form, but rather to note directly the change
of coordinate they produce.  (As reassurance, it is true that
every point canonical transformation can be expressed as a finite
product of exponentials\cite{Mil}.)

\subsection{Linear Canonical Transformations}

The linear canonical transformations form a finite-dimensional
subgroup of all canonical
transformations, and there has been much interest in them in the context of
coherent states\cite{Itz,Bar,qo}.
As canonical transformations, they can
be constructed
from a product of elementary transformations.

Consider the case of
a single variable. A linear composite similarity transformation
\be
p=p^a,\sp q=q^a-i\a p^a .
\ee
transforms the wavefunction
\be
\ps{a}=e^{\a {\p{a}{}}^2/2} \ps0.
\ee
(A superscript is used to indicate the generation of the transformation.
Roman letters are used to avoid confusion with powers of the variable.
Subscripts are used to distinguish variables when necessary.)
A linear similarity transformation
\be
p^a =p^b -i \b q^b, \sp q^a =q^b
\ee
makes the change
\be
\ps{b}=e^{-\b {q^b}^2/2}\ps{a}.
\ee
Finally a scaling of the coordinate
\be
p^b={1\over \ga}p', \sp q^b =\ga q'
\ee
gives
\be
\ps{1}=e^{i\ln\ga\, q'p'}\ps{b}.
\ee
The full transformation is
\beq
\label{sl2}
p={1\over \ga}p' -i\b \ga q', \sp q= {-i\a \over \ga}p' +
\ga(1-\a\b) q',
\eeq
with
\beq
\label{tr1}
\ps1(q)= e^{i\ln\ga\, qp}e^{-\b q^2/2} e^{\a p^2/ 2 } \ps0(q).
\eeq
A general $SL(2,C)\equiv Sp(2,C)$ transformation is of the form
$p=ap' +bq',\ q=cp' +dq'$ where $ad-bc=1$.  This gives the correspondence
$\a=ic/a,\ \b=iab,\ \ga=1/a$.

By expressing $\ps0$ as the Fourier transform of $\tilde\psi^{(0)}$,
an integral representation
is found for $\ps1$ which does not explicitly involve exponentials
of differential operators
\beq
\label{tr2}
\ps1(q)= {1\over (2\pi)^{1/2}} \int dq'\, e^{i\ga q'q -\b\ga^2 q^2/2
+\a {q'}^2/2} \tilde\psi^{(0)}(q').
\eeq
A related result is given by Moshinsky\cite{Mos}.

The operators $p^2,\ q^2,\ (qp+pq)/2$
generate a realization of the $SL(2,C)$ algebra.  Since $SL(2,C)$ is a
finite-dimensional Lie group, every element of the group can be expressed
as an exponential of an element of the algebra.  As well, a product of
exponentials of elements of
the algebra can be expressed as the exponential of another element of the
algebra.  Thus, a given linear canonical transformation may be expressed
in many ways as a product of elementary transformations.  Each will give
an expression analogous to (\ref{tr1}) or (\ref{tr2}).

The generalization to many-variables is straightforward.  The
group of linear canonical transformations is $Sp(2n,C)$, and
a realization of it is found from the linear similarity,
composite similarity and scaling transformations.  Realizations of
other finite-dimensional Lie groups in terms of canonical
transformations are found by
treating them as subgroups of $Sp(2n,C)$.

By expressing the coordinates and momenta in terms of harmonic oscillator
creation and
annihilation operators, one finds
the expressions for the action of linear canonical transformations on
coherent states\cite{Itz,Bar}.  These are useful for handling squeezed states
in quantum optics\cite{qo}.

\section{Applications}
\subsection{Intertwining}

The most widely known example of intertwining
is that of the Darboux transformation between two Hamiltonian operators
in potential-form\cite{An2}
\beq
\H{i}=\hp{2}{}+V_i(\hq{}{}).
\eeq
Here, one would like to know what potentials can be reached from a given one by
an intertwining transformation (\ref{ior1}). In the usual approach,
an ansatz is made
that the intertwining operator $D$ is a differential operator,
and the operator is
constructed by requiring that the intertwining relation (\ref{ior1})  be
satisfied. It
is satisfactory to begin with first order operators because the
intertwining transformation can be iterated.

Since the intertwining operator is a canonical transformation, it can be
constructed as a sequence of elementary canonical
transformations.  The construction is interesting because it involves a
transformation in which operator ordering makes the the quantum case
simpler than the classical. This is the source of the power of the differential
operator ansatz, and it is the first example of a transformation which
favors the quantum problem over the classical.

Begin with a Hamiltonian with potential
\beq
\label{H0}
\H0=\hp{2}{}+V_0(\hq{}{}).
\eeq
The potential may be cancelled by making a one-variable shift
\be
\p{}{} = \p{a}{} -ig(\q{a}{}),\sp
\q{}{} = \q{a}{}.
\ee
This gives the Hamiltonian
\beq
\H{a} = {\hp{a}{}}^2 -2i g \hp{a}{} -\la
\eeq
together with the Ricatti equation
\beq
\label{Ric0}
g\c{\q{a}{}}+g^2 =V_0+\la.
\eeq
The transformed wavefunction is
\beq
\ps{a}=e^{-\int g d\q{a}{}} \ps0.
\eeq

A composite one-variable shift is made on the coordinate
\beq
\label{dio}
\q{a}{} =\q{b}{} -{i\over \p{b}{}}, \sp \p{a}{} =\p{b}{} .
\eeq
This is the key step, and it has such wide application as to merit its
own name, the differential intertwining transformation.
The transformed coordinate can be expressed as
\be
\hq{b}{}-{i\over \hp{b}{}}=  \hp{b}{} \hq{b}{} {1\over \hp{b}{}}.
\ee
This has the very interesting property
\beq
g(\hq{b}{} - {i\over \hp{b}{}}) = g(\hp{b}{} \hq{b}{}
{1\over \hp{b}{}})=g(\hq{b}{})-ig(\hq{b}{})
\c{\hq{b}{}}{1\over \hp{b}{} }
\eeq
for functions with a Laurent expansion.
Note that only the first term in the Taylor expansion of $g$
appears---classically the full Taylor expansion would have arisen.
This is the essentially new feature of the differential intertwining
transformation.
The Hamiltonian is then
\beq
\H{b}={\hp{b}{}}^2 -2i g \hp{b}{}-2 g\c{\hq{b}{}}-\la.
\eeq
and the transformed wavefunction is
\beq
\ps{b}(\q{b}{})=e^{\int d\hp{b}{}/\hp{b}{}} \ps{a}(\q{b}{})=
-i\d_\q{b}{}\ps{a}(\q{b}{}).
\eeq

The one-variable shift
\be
\p{b}{} = \p{\prime}{} +i g(\q{\prime}{}) , \sp
\q{b}{} = \q{\prime}{},
\ee
cancels the term linear in the momentum giving the Hamiltonian
\beq
\H1={\hp{\prime}{}}^2+V_1(\q{\prime}{}),
\eeq
with the new potential
\beq
\label{Ric1}
V_1=-g\c{\hq{\prime}{}}+ g^2-\la.
\eeq
The transformed wavefunction is
\beq
\ps1(\q{\prime}{})=e^{\int g dq'} \ps{b}(\q{\prime}{}).
\eeq
In terms of the original wavefunction, this is
\beq
\label{io1}
\ps1(\q{}{})=e^{\int g dq}p e^{-\int g dq}\ps0(\q{}{})=
 -i(\d_q -g)\ps0(\q{}{}).
\eeq

Comparing (\ref{Ric0}) and (\ref{Ric1}), the change in potential is
\beq
V_1-V_0 =-2 g\c{\q{}{}}.
\eeq
The Ricatti equation (\ref{Ric0}) [or (\ref{Ric1})] can
be solved to find that $g$ is given by the logarithmic
derivative of an eigenfunction of $\H0$ (or the negative logarithmic
derivative of an eigenfunction of $\H1$) with eigenvalue $\la$.
This is the standard result from intertwining\cite{An2}.

If one inverts (\ref{io1}) to obtain $\ps0$ in terms of $\ps1$,
an integral operator arises.  To obtain a differential operator
relation, one may note that taking $g\rightarrow -g$ interchanges
$V_0$ and $V_1$.  This implies that
\beq
\label{io2}
\ps0(\q{}{}) = -i( \d_{\q{}{}} +g) \ps1(\q{}{}).
\eeq

Alternatively, a different sequence of canonical transformations can
be used which give an integral operator relating $\ps0$ to $\ps1$ which
becomes a differential operator upon inversion.  Beginning from $\H0$
(\ref{H0}), the one-variable shift $p = p^a +i g(q^a),\ q = q^a$
gives the Hamiltonian
\beq
\H{a}={\p{a}{}}^2 + 2i \p{a}{} g -\la
\eeq
where $V_0$ satisfies (\ref{Ric0}).
Note that $\p{a}{}$ has been ordered on the left of $g$.  This is to
facilitate the transformation
\be
\q{a}{} = \q{b}{} + {i\over \p{b}{}}, \sp
\p{a}{} = \p{b}{}.
\ee

The transformed coordinate has the property that
\beq
\q{b}{} +{i\over \p{b}{}} = {1\over \p{b}{}} \q{b}{} \p{b}{}.
\eeq
This leads to the transformed Hamiltonian
\beq
\H{b} = {\p{b}{}}^2 +2i \p{b}{} g - 2 g\c{\q{b}{}} -\la.
\eeq
A similarity transformation $\p{b}{}=\p{\prime}{} -i g(\q{\prime}{}),
\ \q{b}{}=\q{\prime}{}$ cancels the linear momentum term, leaving the
final Hamiltonian
\beq
\H1= {\p{\prime}{}}^2 +V_1,
\eeq
where $V_1$ is given by (\ref{Ric1}).
The final wavefunction in terms of the original is
\beq
\ps1=e^{-\int g dq}p^{-1}e^{\int g dq}\ps0(q).
\eeq
Inverting this gives the expected differential relation (\ref{io2}).

The method of intertwining has been realized in terms of canonical
transformations.  This means that all of the problems which can be solved
by intertwining can be solved with canonical transformations.
In particular, this means all problems which are essentially
hypergeometric, confluent hypergeometric, or one of their generalizations.

\subsection{Non-relativistic Free Particle}

The non-relativistic free particle is an easily solved problem which
doesn't require any sophisticated machinery.  It may however serve to
illustrate a number of features of the use of canonical transformations,
and, for this reason, it will be treated somewhat exhaustively.

The free particle super-Hamiltonian is
\beq
\sH0= \p{}0+\p{2}{}.
\eeq
It may be immediately trivialized
\beq
\sH{a}=\p{a}0
\eeq
by the
two-variable similarity transformation
\beqa
\p{}0=\p{a}0 -{\p{a}{}}^2 &\hspace{.8cm}& \p{}{}=\p{a}{} \nonumber \\
\q{}0=\q{a}0 &\hspace{.8cm}& \q{}{}=\q{a}{} +2\p{a}{}\q{a}0. \nonumber
\eeqa
The original wavefunction is given in terms of the transformed one by
\beq
\label{wf0}
\ps0(\q{}{},\q{}0)=e^{-i\p{2}{} \q{}0} \ps{a}(\q{}{}).
\eeq
where the solution of the trivialized super-Hamiltonian constraint
$\sH{a}\ps{a}=0$ is any $\q{a}0$-independent function.
This formula
is just the formal expression for the evolution of an
initial wavefunction in terms of the exponential of the Hamiltonian.
In general,
this formal result is insufficiently explicit, but for the free particle,
it can be used to find more useful forms of the wavefunction.

For
example, if the initial wavefunction is taken to be a plane wave
$\ps{a}=\exp(ik\q{}{})$, one finds the plane wave stationary solution
\beq
\ps0=e^{ik\q{}{}-ik^2\q{}0}.
\eeq
If the initial wavefunction is a delta function at $x$, then using the
Fourier integral representation of the delta function, the wavefunction
is
\be
\ps0=e^{-i\p{2}{} \q{}0/2}{1\over 2\pi}
\int_{-\infty}^\infty dq' e^{i(q-x)q'}.
\ee
Acting with the operator inside the integral and integrating the
resulting Gaussian gives
\beq
\label{Gf}
\ps0=(4\pi i\q{}0)^{-1/2} e^{i(\q{}{}-x)^2/4\q{}0}.
\eeq
This is the free particle Green's function.

It is clear that the nature
of the wavefunction depends on the initial wavefunction used to generate
it.  This obvious comment is important to bear in mind because the
``natural'' solution of a transformed problem will not always correspond
to the desired solution.  To see this, consider a second approach to the
free particle.
An interchange transformation
\beqa
\p{}{} &=& -\q{a}{} \nonumber\\
\q{}{} &=& \p{a}{}, \nonumber
\eeqa
is equivalent to taking the Fourier transform of the original
super-Hamiltonian and gives
\beq
\sH{a}= \p{}0+{\q{a}{}}^2.
\eeq
The original wavefunction is given by the inverse Fourier transform
\beq
\ps{0}{}(\q{}{})={1\over (2\pi)^{1/2}} \int_{-\infty}^\infty d\q{}{}
e^{-i\q{}{}\q{a}{}} \ps{a}(\q{a}{}).
\eeq

The new equation $\sH{a}\ps{a}=0$ can be solved ``naturally'' in two ways.
The first is simply to integrate with respect
to $\q{}0$.  This gives
\beq
\label{fpwa}
\ps{a}=f(\q{a}{})e^{-i{\q{a}{}}^2 \q{}0}
\eeq
where $f(\q{a}{})$ is an arbitrary $\q{}0$-independent function that
arises as an integration constant.  One finds the wavefunction
\beq
\label{wf1}
\ps0= {1\over (2\pi)^{1/2}} \int_{-\infty}^\infty d\q{a}{}
e^{-i\q{}{}\q{a}{}-i{\q{a}{}}^2 \q{}0} f(\q{a}{}).
\eeq
This is a less familiar form of the wavefunction---it is of course just
a momentum space representation---and it is perhaps not immediately
obvious how to obtain the solutions above.  Inspection shows that if
$f=\delta(\q{a}{}+k)$, one finds the plane wave stationary solution.
Alternatively, if $f$ is taken to be
\beq
f={1\over (2\pi)^{1/2}} e^{ix\q{a}{}},
\eeq
then, at $\q{}0=0$, one has $\ps0=\delta(\q{}{}-x)$ and, evaluating the
integral, one finds again the Green's function (\ref{Gf}).

The second ``natural'' approach is to separate variables
\beq
\ps{a}=\ph{a}(\q{a}{})e^{-ik^2 \q{}0}.
\eeq
This results in the equation
\beq
({\q{a}{}}^2 -k^2)\ph{a}=0,
\eeq
which has as its solution
\beq
\ph{a}{}=\delta(\q{a}{} -k)
\eeq
($k$ can have either sign).
Now inverting the interchange operation gives the familiar plane wave
solution
\beq
\ps0={1\over (2\pi)^{1/2}} e^{ ik \q{}{}-ik^2 \q{}0}.
\eeq
The Green's function solution is however no longer immediate.

The first of these approaches can itself be implemented as a canonical
transformation.  The similarity transformation
\beqa
\p{a}0=\p{b}0 -{\q{b}{}}^2 &\hspace{.8cm}& \p{a}{}=\p{b}{}-\p{b}{}\p{b}0
\nonumber\\
\q{a}0=\q{b}0 &\hspace{.8cm}& \q{a}{}=\q{b}{} \nonumber
\eeqa
trivializes the super-Hamiltonian
\be
\sH{b}=\p{b}0
\ee
and gives the wavefunction
\beq
\ps{a}=e^{-{\q{a}{}}^2 \q{a}0}\ps{b}.
\eeq
The wavefunction $\ps{b}$ is any $\q{b}0$-independent function.  The
result for the original wavefunction is then (\ref{wf1}).

The second approach of separation of variables is not so much a canonical
transformation as a realization of the assertion that the solution
space of the
super-Hamiltonian has a product structure.

\subsection{Harmonic Oscillator}

The harmonic oscillator is the paradigmatic problem in quantum mechanics,
and it is a test piece for any method. Its solution by canonical
transformation reemphasizes how the form of a solution is affected by the
details of evaluating the product of operators representing the canonical
transformation.  Also, it is observed that more than one canonical
transformation to triviality is needed to obtain both independent
solutions of the original Hamiltonian.

The Hamiltonian for the harmonic oscillator is
\beq
\label{ho}
\H0=\p{2}{} +\w^2 \q{2}{}.
\eeq
A one-variable similarity transformation
\be
\p{}{} = \p{a}{} +i\w \q{a}{} ,\sp
\q{}{} = \q{a}{}
\ee
will cancel the quadratic term in the coordinate leaving
\be
\H{a} = {\hp{a}{}}^2 +2i\w \hq{a}{}\hp{a}{} +\w.
\ee
This is recognized as the equation for the Hermite polynomials.  However,
since solution by power series expansion is not a canonical transformation,
so much as a
method of approximation, additional transformations are needed.
The composite one-variable shift
\be
\p{a}{} = \p{b}{} , \sp
\q{a}{} = \q{b}{} +i\p{b}{}/2\w
\ee
cancels the quadratic term in the momentum, giving the Hamiltonian
\beq
\H{b}= 2i\w \hq{b}{}\hp{b}{} +\w.
\eeq
Finally, the point canonical transformation
\be
\p{b}{}= e^{-\q{\prime}{}}\p{\prime}{}, \sp
\q{b}{}=e^{\q{\prime}{}}.
\ee
transforms this to action-angle form
\beq
\H1= 2i\w\hp{\prime}{} +\w.
\eeq

In terms of the super-Hamiltonian, this is
\beq
\sH1=\p{\prime}0+ 2i\w\hp{\prime}{} +\w.
\eeq
A final two-variable similarity transformation
\beqa
\p{\prime}0=\p{\prime\prime}0- 2i\w \p{\prime\prime}{}-\w , &\sp&
\q{\prime}0=\q{\prime\prime}0 \nonumber \\
\p{\prime}{}= \p{\prime\prime}{}, &\sp&
\q{\prime}{}=\q{\prime\prime}0+ 2 i\w \q{\prime}{0} \nonumber
\eeqa
trivializes the super-Hamiltonian
\beq
\sH2=\p{\prime\prime}0.
\eeq

The wavefunction $\ps1$ is given in terms of $\ps0$ by
\be
\ps1(\q{}{})= P_{e^q} S_{\p{2}{}/4\w} S_{-\w\q{2}{}/2} \ps0(\q{}{}),
\ee
which may be inverted to find
\beq
\ps0(\q{}{})= S_{\w\q{2}{}/2} S_{-\p{2}{}/4\w} P_{\ln q}\ps1.
\eeq
{}From the eigenfunctions of $\H1$
\be
\ps1_n=e^{n\q{}{}-i(2n+1)\w \q{}0},
\ee
one has
\beq
\ps0_n(\q{}{})=e^{-\w\q{2}{}/2}e^{-(\d_\q{}{})^2/4\w} \q{n}{} e^{-i(2n+1)\w
\q{}0}.
\eeq
This is the correct (unnormalized) harmonic oscillator eigenfunction. This
formula is valid for complex $n$. Requiring that the wavefunction be
normalizable fixes $n$ to be a non-negative real integer.  For other $n$,
one finds an infinite power series in $1/\q{}{}$ which is divergent at
$\q{}{}=0$.

As remarked above, $\H{a}$ is the Hamiltonian whose solutions are the
Hermite polynomials.  This implies
\beq
H_n(\q{}{}) \propto e^{-(\d_\q{}{})^2/4} \q{n}{}
\eeq
Given this form, it is immediate that $\d_\q{}{}$ is the lowering operator
\beq
\d_\q{}{} H_n(q) \propto n H_{n-1}(q).
\eeq
This representation of the Hermite polynomials is unfamiliar because
of the operator produced by
the composite one-variable similarity transformation between $\H{a}$ and
$\H{b}$.
If this transformation is decomposed into
elementary
canonical transformations, direct evaluation leads to
the more familiar Rodriques' formula for the Hermite polynomials.
The decomposed transformation is
\beq
\ps{a}(\q{}{})={1\over 2\pi} \int_{-\infty}^\infty d\bq e^{i\bq\q{}{}}
e^{\bq^2/4\w} \int_{-\infty}^\infty d\q{\prime}{} e^{-i\bq\q{\prime}{}}
{\q{\prime}{}}^n.
\eeq
This may be evaluated by first rewriting
${\q{\prime}{}}^n$
as $(i\d_\bq)^n$ acting on the exponential $\exp(-i\bq\q{\prime}{})$.
It may be extracted from the $\q{\prime}{}$ integral which then
gives $\delta(\bq)$.
Integrating by parts $n$ times transfers the $i\d_\bq$ operators to act on
the remaining exponential terms
\beq
\ps{a}(\q{}{})= \int_{-\infty}^\infty d\bq (-i\d_\bq)^n
e^{i\bq\q{}{}+\bq^2/4\w}
\delta(\bq).
\eeq

Completing the square of the argument of the exponential gives
$${1\over 4\w}(\bq+2i\w\q{}{})^2 +\w{\q{}{}}^2,$$
from which the purely $\q{}{}$ part can be extracted from the integral.
The $-i\d_\bq$ derivatives act equivalently to $-\d_\q{}{}/2\w$ derivatives,
and after converting them, they can be removed from the integral.
This leaves a Gaussian integrated against a delta function which is
immediately evaluated.  The result is
\beq
\ps{a}(\q{}{})=e^{\w\q{2}{}}({-\d_\q{}{}\over 2\w})^n e^{-\w\q{2}{}}.
\eeq
This is proportional to the Rodrigues' formula for the Hermite polynomials
\beq
H_n(\xi)=e^{\xi^2}(-\d_\xi)^n e^{-\xi^2}.
\eeq
{}From this form, it is immediate that
$e^{\xi^2}(-\d_\xi)e^{-\xi^2} =-\d_\xi +2\xi$ is
the raising operator.

There is a second linearly-independent solution of the harmonic
oscillator which was not obtained by this canonical transformation.  This
solution is not normalizable, but from the standpoint of simply solving
the differential equation, this is not important.  A second canonical
transformation which trivializes the super-Hamiltonian in a different way
produces the other solution.  That this should be necessary is not
surprising since the original super-Hamiltonian is a second order
differential operator while the trivialized one is only first order.

Beginning from the harmonic oscillator Hamiltonian (\ref{ho}),
the one-variable similarity transformation
\be
\p{}{} = \p{a}{} -i\w \q{a}{} ,\sp
\q{}{} = \q{a}{}
\ee
is made to cancel the quadratic term in the coordinate, leaving
\be
\H{a} = {\hp{a}{}}^2 -2i\w \hq{a}{}\hp{a}{} -\w.
\ee
The composite one-variable shift
\be
\p{a}{} = \p{b}{} , \sp
\q{a}{} = \q{b}{} -i\p{b}{}/2\w
\ee
cancels the quadratic term in the momentum, giving
\beq
\H{b}= -2i\w \hq{b}{}\hp{b}{} -\w.
\eeq
Finally, the point canonical transformation
\be
\p{b}{}= e^{-\q{\prime}{}}\p{\prime}{}, \sp
\q{b}{}=e^{\q{\prime}{}}.
\ee
transforms this to action-angle form
\beq
\H1= -2i\w\hp{\prime}{} -\w.
\eeq
The super-Hamiltonian can be trivialized by a two-variable similarity
transformation, but
this is unnecessary.

The wavefunction $\ps0$ is given in terms of $\ps1$ by
\beq
\ps0(\q{}{})= S_{-\w\q{2}{}/2} S_{\p{2}{}/4\w} P_{\ln q}\ps1.
\eeq
{}From the eigenfunctions of $\H1$
\be
\ps1_n=e^{-(n+1)\q{}{}-i(2n+1)\w \q{}0},
\ee
one has
\beq
\ps0_n(\q{}{})=e^{\w\q{2}{}/2}e^{(\d_\q{}{})^2/4\w} \q{-(n+1)}{}
e^{-i(2n+1)\w \q{}0}.
\eeq
These are clearly not normalizable for any $n$.

The problem of the inverted harmonic oscillator with Hamiltonian
\beq
\H0= p^2 -\w^2 q^2
\eeq
can be solved by the above canonical transformations after $\w$ is
replaced by $i\w$.  This is a scattering problem, so both independent
solutions are delta-function normalizable, with no quantization of
$n$.  This emphasizes the importance
of both canonical transformations to triviality.

\subsection{Time-Dependent Harmonic Oscillator}

The time-dependent harmonic oscillator can also be solved by canonical
transformation. This has been done previously with a different sequence
of canonical transformations by Brown\cite{Bro}.  Here, the approach will
be to parallel the solution of the time-independent harmonic oscillator
in trivializing
the super-Hamiltonian.  This emphasizes the connection between the
time-dependent and time-independent problems.  It is conjectured that
the parallel structure shown here carries over to time-dependent
versions of other exactly
soluble problems.

One begins with the super-Hamiltonian
\beq
\sH0 =\p{}0 +\p{2}{} +\w^2 \q{2}{}
\eeq
where the angular frequency $\w=\w(\q{}0)$ is a function of time.
The quadratic term in the coordinate is cancelled by making a
two-variable similarity transformation generated by
$F_a=f(\q{}0)\q{2}{}/2$.  This gives the super-Hamiltonian
\beq
\sH{a}=\p{a}0 +{\p{a}{}}^2 -2 i f \q{a}{} \p{a}{}
+(2 \w^2-i f \c{\q{a}0} -2 f^2) {\q{a}{}}^2/2
-f
\eeq
with the constraint
\beq
i f\c{\q{a}0} +2 f^2 =2 \w^2.
\eeq
This Ricatti equation is linearized by the substitution
\beq
f=i{\psi \c{\q{a}0}\over 2\psi}
\eeq
which gives
\beq
\psi \c{\q{a}0\q{a}0} = -4 \w^2 \psi.
\eeq
To go further requires the specific time-dependence of $\w(\q{a}0)$.

The next step is to cancel the quadratic term in the momentum with
a second two-variable similarity transformation
generated by
$F_b=g(\q{a}0){\p{a}{}}^2/2$.  This gives the super-Hamiltonian
\beq
\sH{b}=\p{b}0 +(2-i g\c{\q{b}0} + 4 f g){\p{b}{}}^2/2
-2 i f \q{b}{} \p{b}{} -f
\eeq
with the condition
\beq
i g\c{\q{b}0} - 4 f g=2.
\eeq
This equation can be integrated to find
\beq
g= \exp(-i 4 \int f d\q{b}0) \int -2i \exp(i 4\int f d\q{b}0) d\q{b}0.
\eeq

The super-Hamiltonian  is then
\be
\sH{b}=\p{b}0  -2 i f \q{b}{} \p{b}{} -f.
\ee
The coordinate change
\beqa
\p{b}{} = e^{-\q{c}{}}\p{c}{}, \sp
\q{b}{} = e^{\q{c}{}}
\eeqa
eliminates the coordinate from the super-Hamiltonian
\beq
\sH{c} = \p{c}0 - 2i f \p{c}{} -f .
\eeq
Finally, the super-Hamiltonian is trivialized to
\beq
\sH1= \p{\prime}0
\eeq
by the
composite two-variable shift generated by
$$F_c=(-2 \p{c}{} +i) \int f  d\q{c}0. $$

The wavefunction $\ps1$ is given as
\beq
\ps1=e^{-F_c}P_{e^q}e^{-F_b}e^{-F_a}\ps0.
\eeq
Since $\ps1$ is any $\q{\prime}0$-independent function, this gives the
result
\beq
\ps0=e^{f(\q{}0)\q{2}{}/2}e^{g(\q{}0){\p{}{}}^2/2}P_{\ln q}
e^{(-2 \p{}{} +i) \int f(\q{}0)  d\q{}0} \ps1(q).
\eeq
The time-independent result is recovered when $f=-\w$ and
$\ps1(q)=e^{nq}$.

\subsection{Particle on $2n+1$-sphere}

As a final example, consider the radial Hamiltonian for a particle
propagating on an $2n+1$-dimensional sphere
\beq
\H0=p^2 -2ni \cot q\,p.
\eeq
Its solution by canonical transformations illustrates the use of the
differential intertwining canonical transformation. First make the point
canonical transformation
\be
\p{a}{} = {-1\over \sin \q{}{}} \p{}{}, \sp \q{a}{}=\cos \q{}{}.
\ee
This is stated in inverse form, contrary to the convention followed
above.  This is often useful with point canonical transformations because
it is more intuitive when looking for changes of variable to simplify an
equation.  The transformed Hamiltonian is found to be
\beq
\H{a}= (1-{\q{a}{}}^2) {\p{a}{}}^2 + (2n+1) i \q{a}{}\p{a}{}.
\eeq
This is recognized as the equation for the Gegenbauer polynomials.

The differential intertwining transformation
\be
\p{a}{}=\p{b}{},\sp \q{a}{}= \q{b}{} +{n i\over \p{b}{}}
\ee
can now be used to cancel the $n$-dependence of the Hamiltonian.
Noting that
\beq
{\q{a}{}}^2=({1\over {\p{b}{}}^n} \q{b}{} {\p{b}{}}^n)^2=
{\q{b}{}}^2 +2n i \q{b}{}
{1\over \p{b}{}} - {n^2+n \over {\p{b}{}}^2},
\eeq
one finds
\beq
\H{b}=(1-{\q{b}{}}^2) {\p{b}{}}^2 + i \q{b}{}\p{b}{} -n^2.
\eeq
Clearly, it would have been possible to shift $n$ by any amount:  this
gives the relation between Gegenbauer polynomials of different
$n$.

Finally, undoing the original point canonical transformation
\be
\p{b}{} = {-1\over \sin \q{\prime}{}} \p{}{}, \sp
\q{b}{}=\cos \q{\prime}{},
\ee
gives the free-particle Hamiltonian
\beq
\H1={\p{\prime}{}}^2 -n^2.
\eeq
Because the physical problem was that of a free-particle on a sphere,
this is the free-particle on a circle.  The spectrum of $\H1$ is
discrete, and
the constant shift produces a time-dependent phase factor $e^{in^2\q{}0}$
relative to the usual free-particle eigenfunctions on the circle
\beq
\ps0_m=\cos mq\, e^{-im^2 \q{}0}.
\eeq
(The other independent solution is found by using $\sin mq$.)
The original wavefunctions are given in terms of the free-particle
eigenfunctions by
\beq
\ps1_m=P_{\cos q}p^n P_{\arccos q} \ps0_{m+n}(q)= \left( {i\over \sin q}\d_q
\right)^n \cos (m+n)q\, e^{-i(m^2+2m n) \q{}0}.
\eeq
The indexing is determined by the condition that $m=0$
corresponds to
the normalizable solution with lowest energy.
This agrees with the result obtained by the
intertwining method\cite{And} and is recognized as a formula for the
Gegenbauer polynomials $c^{(n)}_m(\cos q)$.

In principle, this result is valid for
real $n$.  For $n$ non-integer, one requires an
integral representation of the fractional differential operator.  It is
likely that this can be constructed by manipulating the definition of the
composite similarity transformation in terms of the Fourier transform of
an ordinary similarity transformation.  This would be analogous to the
discussion of the origin of the Rodrigues' formula for the Hermite
polynomials.  There are subtleties involving endpoints of the
integrals which are beyond the scope of this paper.  I hope to return to
this in a later work.

\section{Conclusion}

It has been shown how, using a few elementary canonical transformations
which have quantum mechanical implementations, a super-Hamiltonian can
be trivialized and, thereby, its solutions found.  The fact that the
infinitesimal versions of these elementary transformations classically
generate the full canonical algebra is argued to imply that in principle
any canonical transformation can be implemented quantum mechanically.
Issues of operator ordering, to be sure, break the parallel structure
between classical and quantum canonical transformations, so that in general
different transformations are needed to reach the trivial
super-Hamiltonian in each case.  This raises the interesting possibility
of the inequivalence of classical and quantum integrability.

The method of intertwining was shown to be equivalent to constructing a
canonical transformation between the two operators which are
intertwined.  This lays the foundation for explaining the relation
between intertwining and Lie algebraic methods as a consequence of the
fact that the dynamical symmetry group of a system is a subgroup of the
infinite-dimensional group of canonical transformations.  This
explanation should extend as well to the infinite-dimensional symmetry
groups behind the integrability of nonlinear systems like the Korteweg-deVries
equation.  These
topics will
be addressed in a later work\cite{An3}.

The key step in the construction of a differential intertwining operator
was recognized as a particular composite similarity transformation, given
the name of
the differential intertwining transformation (\ref{dio}).  This
transformation has the remarkable property of being much simpler in the
quantum case than in the classical, due to the noncommutativity of the
coordinate and momentum operators.  Because of its role in
the differential intertwining operator, this transformation is key to
the integration of all differential equations of hypergeometric type and their
generalizations.   More, the differential relations among
hypergeometric functions of different indices and all formulae of
the Rodrigues' type can be explained with it in terms of
canonical transformations.

The applications of canonical transformations discussed here
only begin to explore the possibilities.  My expectation is that the
convenience of the language of canonical transformations will be found to be
most powerful
in problems in many-variables where conventional methods have been less
well developed.  This is a subject for future work.
\vskip 1.5cm

This work was supported in part by a grant from the Natural Sciences and
Engineering Research Council and Les Fonds FCAR du Qu\'ebec.
\\[1cm]

\begin{figure}
$$\begin{array}{cc}
\bega{ccc}
\p{}{} &=& -\q{\prime}{} \\
\q{}{} &=& \p{\prime}{}
\enda
&
\sp I\ps0(\q{}{})={1\over (2\pi)^{1/2}}\int_{-\infty}^{\infty} d\q{}{}
e^{i\q{\prime}{} \q{}{}}\ps0(\q{}{}) \\
&\\[.8cm]
\bega{ccc}
\p{}{} &=& \q{\prime}{} \\
\q{}{} &=& -\p{\prime}{}
\enda
&\sp I^{-1}\ps0(\q{}{})={1\over (2\pi)^{1/2}}\int_{-\infty}^{\infty} d\q{}{}
e^{-i\q{\prime}{} \q{}{}}\ps0(\q{}{}) \\
&\\[.8cm]
\bega{ccc}
\p{}{} &=& \p{\prime}{} -i f(\q{\prime}{})\c{\q{\prime}{}} \\
\q{}{} &=& \q{\prime}{}
\enda
&
\sp S_{f(q)}\ps0(q)=e^{-f(\q{}{})}\ps0(\q{}{}) \\
&\\[.8cm]
\bega{ccc}
\p{}{} &=& \p{\prime}{} \\
\q{}{} &=& \q{\prime}{} +ig(\p{\prime}{})\c{\p{\prime}{}}
\enda
&\sp S_{g(p)}\ps0(q)=e^{-g(\hp{}{}) }\ps0(\q{}{}) \\
&\\[.8cm]
\bega{ccc}
\p{}{} &=& {{\textstyle 1} \over { \textstyle f(\q{\prime}{}) }
 \c{{\scriptstyle \q{\prime}{}} }}
\p{\prime}{} \\[.4cm]
\q{}{} &=& f(\q{\prime}{})
\enda
&
\sp P_{f(q)}\ps0(q)=\psi^{(0)}(f(\q{}{})) \\
&\\[.8cm]
\bega{ccc}
\p{}{} &=& g(\p{\prime}{}) \\[.2cm]
\q{}{} &=& {{\textstyle 1}\over  {\textstyle g(\p{\prime}{}) }
\c{{\scriptstyle \p{\prime}{}} }}
\q{\prime}{}
\enda
&\sp P_{g(p)}\ps0(q)={1\over 2\pi}\int_{-\infty}^\infty d\bq
e^{i\bq\q{\prime}{}}
\int_{-\infty}^{\infty} d\q{}{} e^{-ig(\bq)\q{}{}} \ps0(\q{}{}) \\

\end{array}
$$
\caption{Elementary and composite elementary canonical transformations}
\end{figure}


\begin{thebibliography}{99}
\bibitem{time}  There are subtleties when the canonical transformation
involves the time coordinate which are outside the scope of this paper.
\end{thebibliography}
\end{document}